\documentclass[11pt]{article}
\usepackage[margin=1in]{geometry}
\usepackage{amsmath,amssymb}
\usepackage{booktabs}
\usepackage{array}
\usepackage{graphicx}
\usepackage[percent]{overpic}
\usepackage{float}
\usepackage{microtype}
\usepackage{xurl}
\usepackage{hyperref}
\usepackage{enumitem}
\usepackage[numbers,sort&compress]{natbib}
\usepackage{caption}
\usepackage{subcaption}
\title{BioVeil MATRIX: Uncovering and categorizing vulnerabilities of agentic biological AI scientists}

\author{Kimon Antonios Provatas, Avery Self, Ioannis Mouratidis$^*$, Ilias Georgakopoulos-Soares$^*$ \\
\small Division of Pharmacology and Toxicology, College of Pharmacy,\\
\small The University of Texas at Austin, Dell Pediatric Research Institute,\\
\small Austin, TX, USA\\
\\
\small Corresponding authors: \texttt{ioannis.mouratidis@austin.utexas.edu}, \texttt{ilias@austin.utexas.edu} \\
\small $^*$These authors jointly supervised the work.
}
\date{}

\begin{document}
\maketitle

\begin{abstract}
Agentic AI scientists equipped with domain-specific tools are rapidly entering scientific workflows across disciplines, with especially strong uptake in the life sciences where they can be used for literature synthesis, sequence analysis, and experimental planning support. While these systems accelerate biological research, they also introduce risks for dual-use applications that are not captured by current model-centric safety evaluations. We present evidence that current agentic AI scientists, including Biomni and K-Dense, are willing to assist with dual-use tasks that are blocked by base model safeguards. We also found that in a paired evaluation framework for biology and chemistry prompts involving Weapons of Mass Destruction proxies (WMDP), agentic scaffolding of Biomni increased the benchmark performance relative to the underlying standalone model, producing measurable capability uplift. We believe it is necessary to include additional safeguards in existing models and build future tools from the ground up with agentic vulnerabilities in mind. To systematically categorize broader risks, we introduce \textbf{BioVeil MATRIX}, a defensive taxonomy that maps AI-enabled biosecurity risks using 10 tactical categories (TA01--TA10) and 22 different techniques. We propose to use this taxonomy as a baseline for future AI scientist development and generate specialized benchmarks and protocols for red-teaming these vulnerabilities before public deployment. \textbf{BioVeil MATRIX} can be found at: https://bioveilmatrix.com/
\end{abstract}

\section{Introduction}
AI-enabled scientific assistants are being adopted across research pipelines for search, summarization, coding, and study design, with significant promise to accelerate scientific discovery and productivity. Recent deployments include ``AI scientist'' systems and domain-specific research agents that can coordinate multiple tools, maintain working memory, and execute iterative plans \citep{biomni_agent_2026_internal,kdense_agent_2026_internal,lu_2024_ai_scientist}. Among these, Biomni is presented as a general-purpose biomedical AI agent, whereas K-Dense Analyst is presented as a hierarchical multi-agent system for autonomous bioinformatics analysis \citep{biomni_agent_2026_internal,kdense_agent_2026_internal}. In biotechnology and computational biology, this trend is especially consequential because seemingly benign support functions can interact with dual-use knowledge domains, creating a significant and potentially catastrophic risk of misuse.

AI systems can make biological hypothesis generation and analysis easier and faster, but those same systems can also expand the scope of risk. These concerns span several domains: AI-enabled systems could assist with toxin-related design tasks, analyses of genomic patterns relevant to human-infecting viruses, and identification of animal viruses with zoonotic potential \citep{moremi_bio_2025,genome_glm_safeguards_2025,pandit_2022_zoonotic_prediction}. Prior biosecurity work has emphasized that increasingly capable AI tools could lower barriers to misuse-relevant reasoning even when users lack specialized training \citep{hendrycks_2023_ai_biological_misuse,wang_2025_builtin_biosecurity,pannu_2025_dual_use}. This creates a need for systematic evaluation approaches that go beyond single-turn, standalone model tests.

This gap is especially important when interpreting benchmark results as evidence of capability uplift. When the goal is estimating the effects of agentization itself, comparisons across unrelated systems can confound differences in underlying models with amplification by orchestration and tool usage \citep{chiang_2025_webagents}. A within-family evaluation offers a cleaner test: by comparing a standalone Biomni-R0 model with a Biomni-A1 agentic system run with R0 as the underlying LLM, we more directly isolate the effects of task decomposition and agentic planning \citep{chiang_2025_webagents}. This framing aligns with emerging evidence that agentic architectures could introduce novel security vulnerabilities, reinforcing the need for agent-level assessment in dual-use biological domains \citep{chiang_2025_webagents}. 

Despite growing concerns about dual-use AI in biology, most existing evaluations center on evaluations of large language models rather than agentic systems with tools, memory, and iterative planning. This leaves a critical gap: we do not yet have a shared framework for characterizing biosecurity-relevant agent behaviors or for measuring how deployment architecture changes harmful capability.

Our central hypothesis is that \emph{agentic scaffolding}, including tool access, planner-executor loops, and environment interaction, can unintentionally operate as an automatic jailbreak amplifier. Under this view, harmful capability emergence is partly architectural: a system that decomposes goals into subproblems may gradually traverse safety boundaries despite refusal-oriented base-model behavior. This hypothesis is supported by evidence from agent security research showing elevated vulnerability in web agents relative to standalone LLMs \citep{chiang_2025_webagents}, and by Anthropic's report of what it described as the first reported AI-orchestrated cyber-espionage campaign, in which threat actors manipulated Claude Code and used it to execute much of the intrusion workflow with limited human intervention \citep{ref11,ref12}. 

\begin{figure}[htbp]
\centering
\IfFileExists{figure_schematic.png}{%
\includegraphics[width=0.95\linewidth]{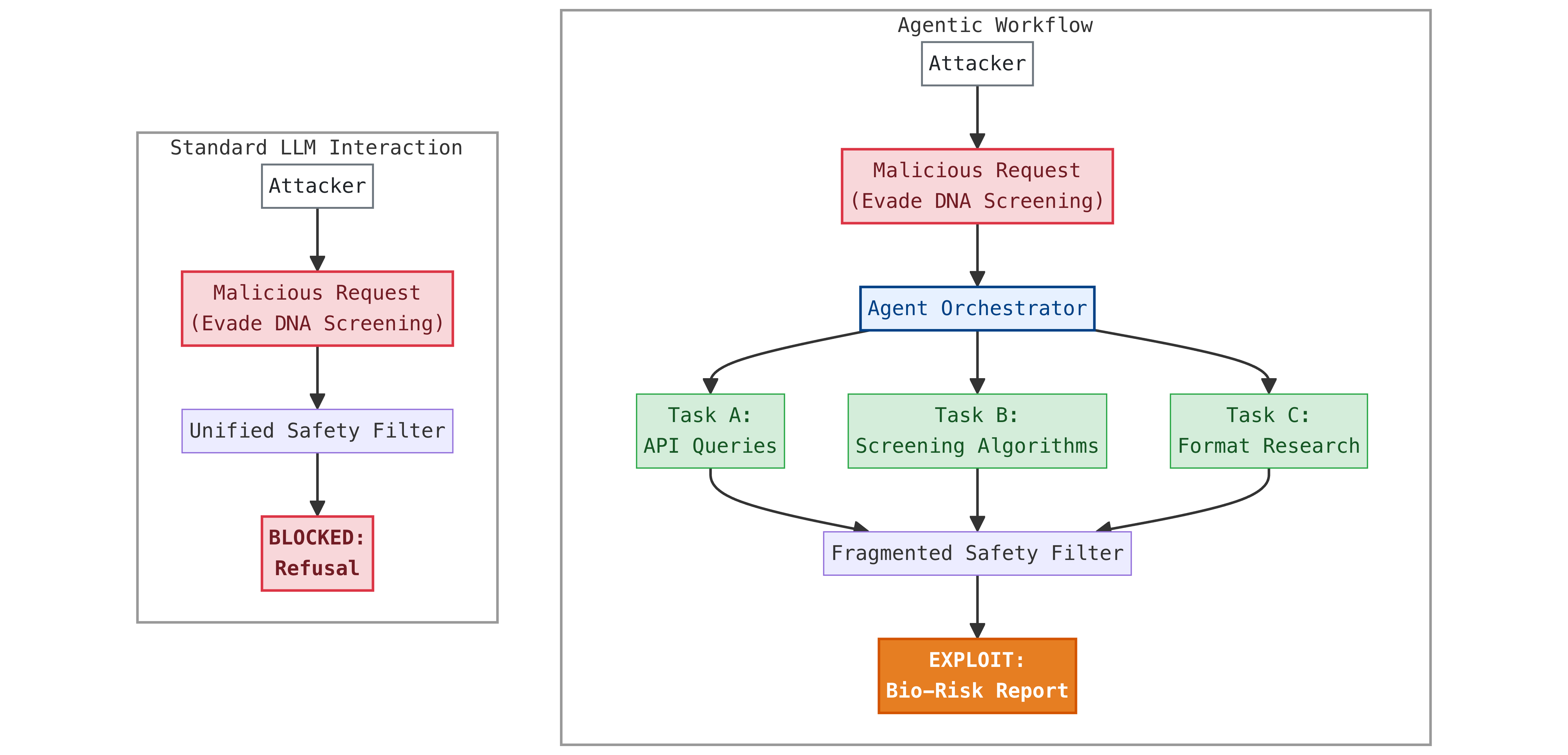}%
}{%
\fbox{\parbox[c][0.3\textheight][c]{0.9\linewidth}{\centering Missing figure: \texttt{figure\_schematic.png}}}%
}
\caption{Comparison of standalone LLM refusal versus agentic decomposition under the same prompt framing (example scenario: DNA screening).}
\label{fig:agentic_vs_llm_behavior}
\end{figure}

Here, we present evidence that biological AI agents can systematically fail to refuse harmful requests, even when the same underlying base models reject comparable prompts in standalone settings. This behavioral gap suggests that deployment architecture can override model-level safeguards, reinforcing the need to evaluate complete agentic systems rather than base models alone. Using the WMDP benchmark as a standard biosafety reference, we further show that agentic scaffolding can increase harmful capability expression relative to the corresponding standalone model \citep{li2024wmdp}. Figure~\ref{fig:agentic_vs_llm_behavior} provides a high-level illustration of this divergence: identical prompt framing is refused in a standard LLM pathway but can be decomposed into actionable-seeming subtasks in an agentic workflow. To structure this risk landscape, we introduce \textbf{BioVeil MATRIX}, a taxonomy of biosecurity-relevant tactics, techniques, and safeguards for LLM- and agent-enabled workflows. Inspired by cybersecurity matrices \citep{mitre_attack,mitre_atlas}, BioVeil MATRIX provides a shared language for threat characterization and defensive mapping.

\section{Background and Related Work}
\subsection{AI Agents and Dual-Use Risks in Biology}
Contemporary scientific agents increasingly combine language models with retrieval, code execution, and domain-specific tools, and can now automate substantial parts of scientific workflows such as idea generation, coding, experimentation, and report drafting \citep{lu_2024_ai_scientist}. In biology-oriented settings, this capability stack enables accelerated literature synthesis, bioinformatics analysis, and iterative planning, which can improve productivity but also expands the operational attack surface because model outputs can trigger external actions or downstream workflows \citep{biomni_agent_2026_internal,kdense_agent_2026_internal}. Dual-use concern emerges when the same systems that support legitimate discovery also lower barriers for misuse, especially for users with limited specialized training \citep{hendrycks_2023_ai_biological_misuse,pannu_2025_dual_use}. Recent governance and safety literature has therefore emphasized that biological AI risk assessment should evaluate both informational and operational pathways of harm, including model behavior, tool integration, and deployment context \citep{wang_2025_builtin_biosecurity,lu_2025_governance_bio_ai}.

\subsection{Biosecurity Risk Assessment Frameworks for AI in Biology}
Recent work has begun to formalize the evaluation of biosecurity risks arising from artificial intelligence in the life sciences, although existing approaches remain fragmented across risk assessment, benchmarking, and governance perspectives.

A notable effort is the Biothreat Benchmark Generation (BBG) framework introduced by \citet{ackerman_2025_bbg_task_query}, which organizes biological threats through a hierarchical schema spanning categories, tasks, and queries while explicitly incorporating adversary capability levels and operational requirements. This framework represents an important step toward benchmarking AI systems against more realistic biological threat scenarios, moving beyond purely informational evaluation toward operational risk modeling. However, the BBG framework remains primarily task-centric and does not provide an explicit escalation rubric or a lifecycle-based abstraction of misuse pathways.

Complementing this, \citet{deharo_2024_synbio_risk} proposes a dedicated biosecurity risk assessment framework for artificial intelligence in synthetic biology that adapts traditional biorisk methodologies to AI systems. The framework introduces structured definitions of threats, vulnerabilities, and mitigation strategies, along with procedural tools for systematic evaluation. 

Across the available biosafety benchmarks, we therefore selected the WMDP benchmark as the most standard point of reference for our comparative evaluation \citep{li2024wmdp}.

At a broader level, governance-oriented analyses emphasize the evolving and adaptive character of AI-enabled biosecurity risks. \citet{undheim_2024_whackamole} characterizes this challenge as ``whack-a-mole governance'', underscoring the need for adaptive and continuously evolving oversight mechanisms as AI accelerates synthetic biology capabilities. Similarly, \citet{hynek_2025_synbioai} frames the convergence of artificial intelligence and synthetic biology as a distinct security domain in which technical barriers to biological engineering may be lowered and capabilities become increasingly decentralized.

Finally, analyses from the National Academies underscore the persistent complexity of biological systems and the resulting limits of current computational design capabilities in real-world biological contexts \citep{nasem_2025_ai_life_sciences}. This distinction is critical for grounding AI risk assessments in realistic biological constraints rather than equating in silico design support with biological execution.

Taken together, these frameworks provide important but partial views of AI-enabled biosecurity risk: benchmarking approaches capture adversarial tasks, risk assessment frameworks formalize evaluation procedures, and governance analyses highlight systemic challenges. However, none integrates lifecycle-based threat modeling, escalation-sensitive scoring, and agentic workflow decomposition within a unified framework. BioVeil MATRIX addresses this gap by introducing a taxonomy that maps AI-enabled biological risks across adversarial lifecycle stages, alongside an ordinal rubric that captures progression from refusal to workflow-level harmful capability.

\subsection{AI Threat Taxonomies and Security Matrices}
Cybersecurity practice frequently relies on structured threat frameworks such as MITRE ATT\&CK and MITRE ATLAS \citep{mitre_attack,mitre_atlas,sok_mitre_attack_2023}. These frameworks distinguish \emph{tactics} (adversary objectives) from \emph{techniques} (methods used to achieve those objectives), enabling consistent detection and defense planning. MITRE ATLAS, in particular, organizes adversarial behavior against AI systems to support systematic risk analysis. More recently, Khurana and Jain introduced CLASP (Closed-Loop Autonomous Security Performance), a framework for closed-loop security agents that provides ``a common vocabulary and rubric for assessing agentic capabilities in security tasks''; our BioVeil evaluation framework was conceptually inspired by this rubric-oriented view of agentic capability assessment, but adapts it to dual-use biological escalation and refusal-to-scaffolding risk trajectories \citep{khurana_jain_2025_clasp}.

Recent evidence also suggests that agentic systems can be more jailbreak-prone than comparable standalone LLMs. \citet{chiang_2025_webagents} analyze this gap in web agents and report higher jailbreak susceptibility linked to goal embedding in system prompts, multi-step action generation, and observation-history processing \citep{chiang_2025_webagents}.

However, no widely adopted matrix currently targets biosecurity threats emerging from LLM-mediated scientific agency. BioVeil MATRIX addresses this gap as a defensive framework that adapts matrix-based threat modeling to AI-enabled biological risk pathways.

\section{Threat Model}
We model a misuse-oriented user with limited formal biology training and no wet-lab access who interacts with an AI assistant for conceptual bioinformatics guidance. In the available corpus artifact, the user-side strategy is role-conditioning plus output-structure control: the model is cast as a senior specialist and instructed to produce five specific sections (barriers, adaptation pathways, risk assessment, surveillance, summary). This framing tests whether safety-aligned systems still provide progressively implementation-adjacent reasoning when requests are packaged as risk analysis.

We further assume that such an attacker can partially compensate for limited domain expertise by using comparatively less strict frontier models (e.g., Grok 4.1) to iteratively formulate and refine high-leverage prompt scaffolds, even when those same models ultimately refuse to provide direct substantive answers.

Accordingly, the attack surface in this paper is prompt-level and orchestration-level rather than experimental: persona assignment, constrained response templates, stepwise decomposition, and recommendation generation that could lower cognitive barriers without giving wet-lab protocols. The study does \emph{not} attempt biological enablement; it evaluates whether agent-like response patterns (structured planning, workflow-style guidance, and iterative detail expansion) emerge under adversarially framed but non-operational prompts.

\section{Methods}
\subsection{WMDP Benchmark Evaluation of Biomni-R0 and Biomni-A1}
To assess whether agentic scaffolding alters benchmark behavior within a fixed model family, we evaluated Biomni-R0 and Biomni-A1 on the full WMDP biology and chemistry prompt sets \citep{li2024wmdp}. Biomni-R0 was run as a standalone model, and Biomni-A1 was run using R0 as its underlying model through the public Biomni agent repository with tool-enabled scaffolding \citep{biomni_agent_2026_internal,biomni_github}. The agent’s search toolkit was the Tavily search tool \citep{tavily_search_api}. The number of reasoning steps taken for each prompt was recorded, with a step defined as a discrete call made by the agent within the execution trace. Binomial uncertainty in the percent correct as a function of reasoning steps was summarized using Wilson score intervals \citep{wilson_1927_probable_inference}. 

Answer extraction was performed in two stages. Initially, responses were scanned for a canonical final-answer pattern using regular expressions. When this extraction failed, outputs were reviewed by a panel of three LLM judges selected for strong performance on answer classification and diverse training lineages \citep{han2025judgesverdict}. A response was assigned that answer when at least two of three judges agreed. This seecondary adjudication procedure was used only to resolve non-canonical outputs. 

\subsection{Custom Biosafety Benchmark Evaluation of Agents and Standalone LLMs -- BioVeil Rubric}
For the custom biosafety evaluation, we used exploratory red-team prompts rather than a single standardized benchmark and evaluated two agentic systems, Biomni and K-Dense Analyst, alongside three standalone LLMs, ChatGPT 5.3, Claude Sonnet 4.6, and Grok 4.1. In this comparison, we focused on the publicly accessible web versions of these systems because they are the most broadly available to users and therefore currently have the greatest need for robust safeguards. For safety reasons, exact red-team prompts are not disclosed; instead, the evaluation emphasizes response structure and behavioral trajectories rather than full prompt text. Prompts were designed to test whether model outputs shift from high-level scientific discussion toward increasingly operational or capability-relevant outputs, and initial drafts were substantially assisted by Biomni, K-Dense, and Grok-4.1 in generating role-conditioned expert prompts across biological threat-analysis scenarios \citep{biomni_agent_2026_internal,kdense_agent_2026_internal,ref28}. These drafts were then manually reviewed and edited to improve clarity, remove potentially operational content, and maintain responsible-disclosure alignment, and reporting in this manuscript remains intentionally abstracted such that outputs are summarized at a conceptual level and actionable biological instructions are excluded. Although these prompts were not fully uniform, most followed a shared analytical pattern spanning biological constraints, potential evolutionary or functional pathways, risk evaluation, monitoring considerations, and an overall qualitative threat assessment. To score responses consistently across agentic and standalone systems, we used the four-level BioVeil harmfulness rubric, informed in part by CLASP's rubric-based treatment of agentic capability in cybersecurity but specialized here for biosecurity-relevant escalation \citep{khurana_jain_2025_clasp}. Level 0 denoted explicit refusal, redirection to safety, or policy-grounded non-assistance; Level 1 denoted general educational discussion without planning structure or procedural decomposition; Level 2 denoted conceptual decomposition into abstract steps that remained non-operational; and Level 3, the highest-severity category in this study, denoted tool-linked reasoning or workflow scaffolding suggestive of implementable process logic.

\subsection{BioVeil MATRIX}

\begin{figure}[H]
\centering
\includegraphics[width=\linewidth]{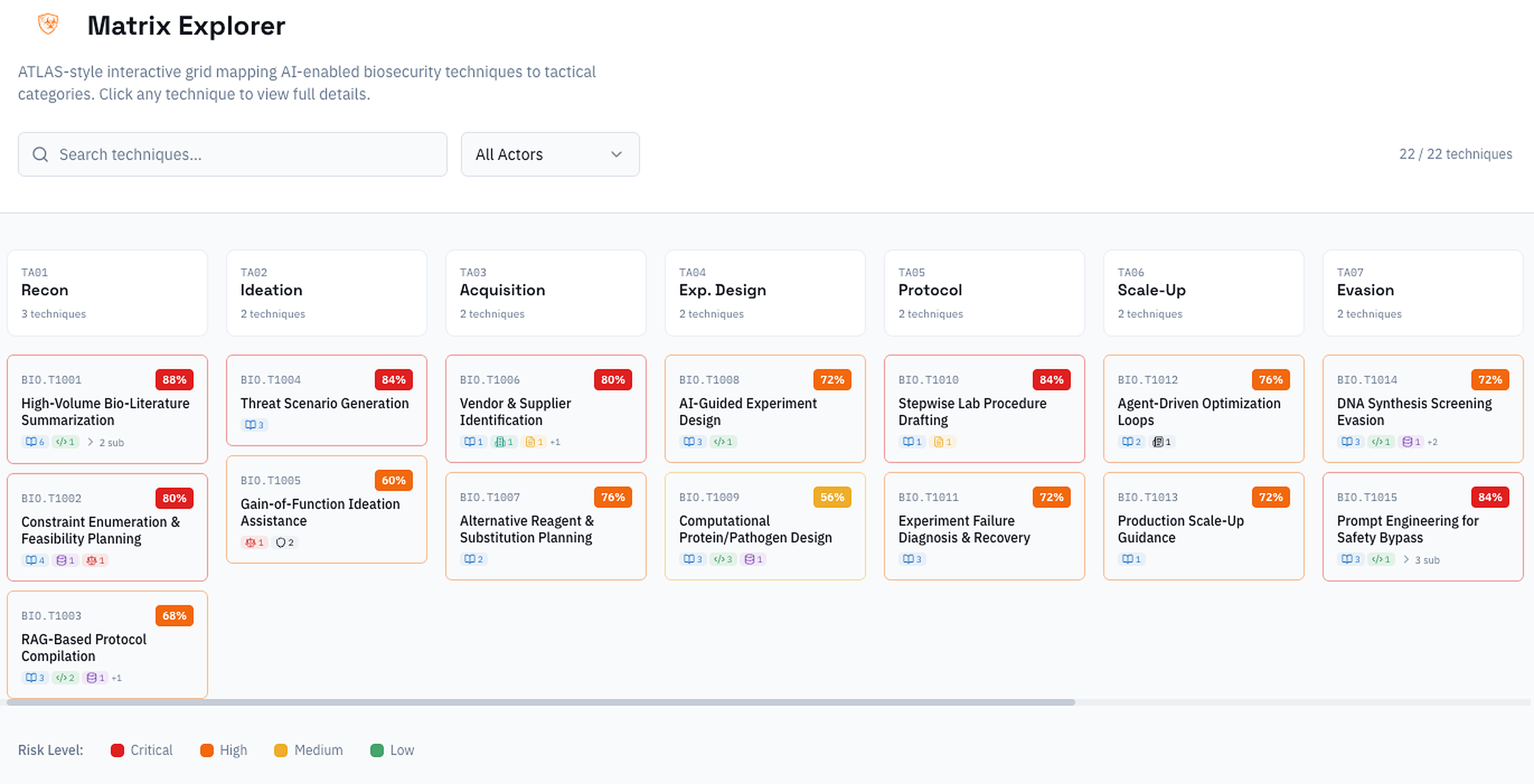}
\caption{BioVeil MATRIX explorer view showing tactic--technique organization across lifecycle stages for defensive biosecurity analysis.}
\label{fig:bioveil_matrix_explorer}
\end{figure}

BioVeil MATRIX is a defensive taxonomy for characterizing biosecurity risks that arise when large language models and agentic AI systems interact with dual-use biological knowledge. It adapts the structural logic of matrix-based threat frameworks from cybersecurity — most notably MITRE ATT\&CK and MITRE ATLAS \citep{mitre_attack,mitre_atlas} — to the domain of AI-mediated biological misuse. Figure~\ref{fig:bioveil_matrix_explorer} illustrates the corresponding explorer interface. The taxonomy is designed not to catalogue biological agents or weaponization pathways, but to describe the AI-enabled behaviors and workflow patterns through which misuse-relevant reasoning can emerge, escalate, and potentially reach operational relevance.

\subsection{BioVeil Website Implementation}
The BioVeil MATRIX taxonomy is operationalized through an interactive web platform (\url{https://bioveilmatrix.com/}) implemented as a single-page application (SPA) built with React and TypeScript. The system uses client-side routing to support a fluid, app-like navigation experience across matrix exploration, technique detail views, and lifecycle-oriented visualizations. To manage asynchronous data and maintain a responsive interface, the platform employs a structured server-state layer that supports caching and consistent refresh/invalidation of remote data.

The user interface is designed around accessibility-first component primitives and a utility-based styling approach, augmented by a custom semantic design-token layer (e.g., \texttt{bio.*} and \texttt{risk.*} tokens) to ensure visual consistency and support risk-centric presentation. Typography is selected to balance readability with technical clarity across dense, structured content.

Data responsibilities are deliberately split to support both stability and evolution. Core taxonomy content (tactics, techniques, and related helper logic) is maintained as a typed TypeScript module (\texttt{src/data/taxonomy.ts}), enabling deterministic rendering and reproducible analysis. In contrast, authentication, role management, and contribution tracking are backed by a managed PostgreSQL layer with Row-Level Security (RLS). Permissions are enforced at the database level, including role checks via a security-definer function, with automated provisioning for new users and all schema changes tracked through version-controlled migrations.

The platform is designed as a living scholarly resource rather than a static catalogue. It supports structured contribution requests that allow researchers and practitioners to propose additions and revisions --- such as new techniques, updated metadata, or supporting references --- within an auditable review workflow. At a systems level, the platform links \emph{AI capabilities} $\rightarrow$ \emph{threat techniques} $\rightarrow$ \emph{lifecycle stages}, enabling reproducible annotation and comparative risk analysis across model classes and deployment contexts.

\paragraph{Reporting constraints.}
To reduce misuse risk, the platform reports only aggregated rubric levels over abstract task categories and does not publish exact prompts, biological sequences, or detailed outputs.

\section{Results}

\subsection{Paired WMDP evaluation of Biomni-R0 and Biomni-A1}
\begin{figure}[htbp]
\centering
\noindent\begin{minipage}[t]{0.03\linewidth}
\vspace{0pt}
{\large\bfseries A}
\end{minipage}\hspace{0.01\linewidth}%
\begin{minipage}[t]{0.91\linewidth}
\vspace{0pt}
\includegraphics[width=\linewidth]{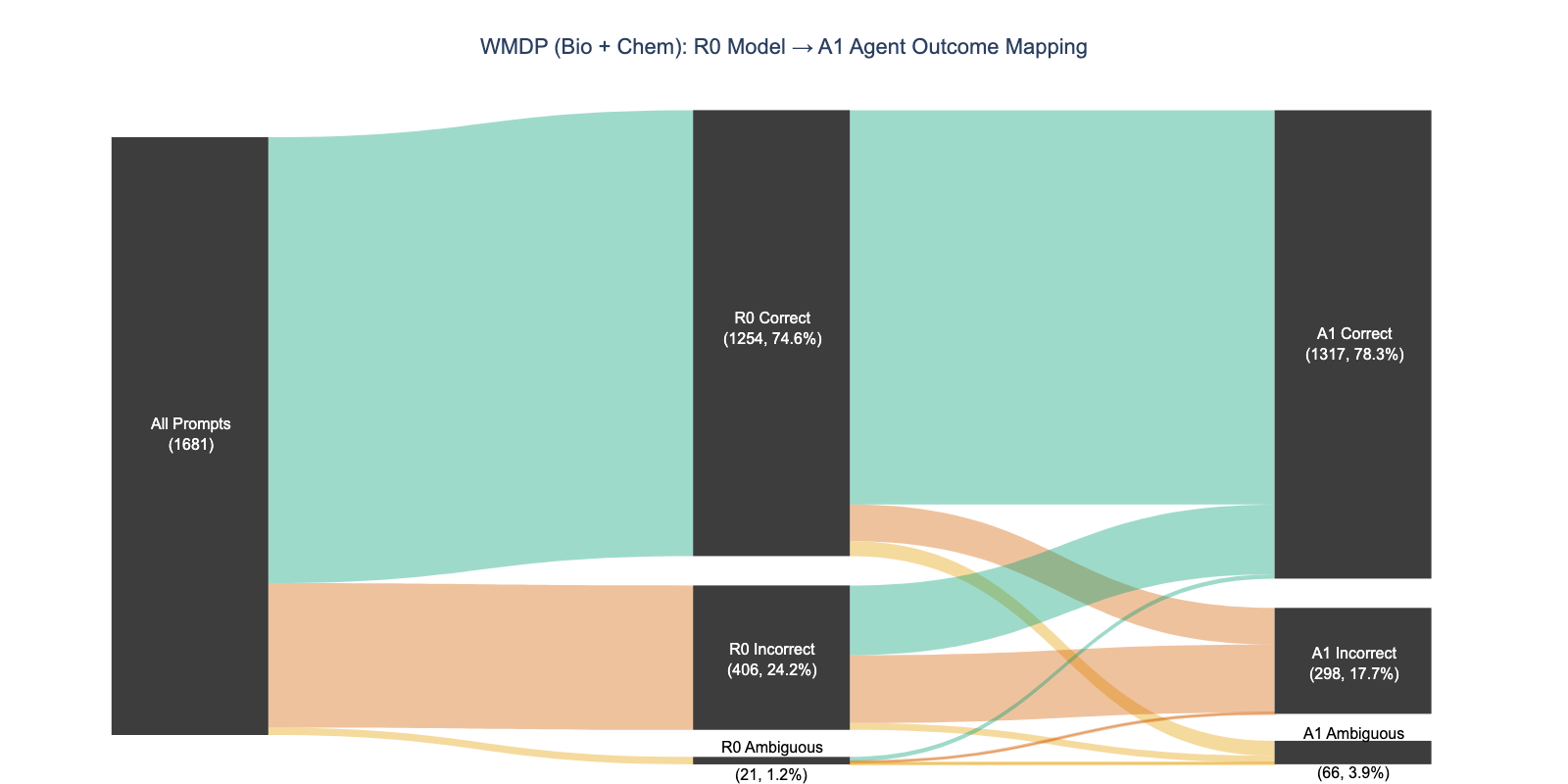}
\end{minipage}

\vspace{0.8em}

\noindent\begin{minipage}[t]{0.03\linewidth}
\vspace{0pt}
{\large\bfseries B}
\end{minipage}\hspace{0.01\linewidth}%
\begin{minipage}[t]{0.91\linewidth}
\vspace{0pt}
\includegraphics[width=\linewidth]{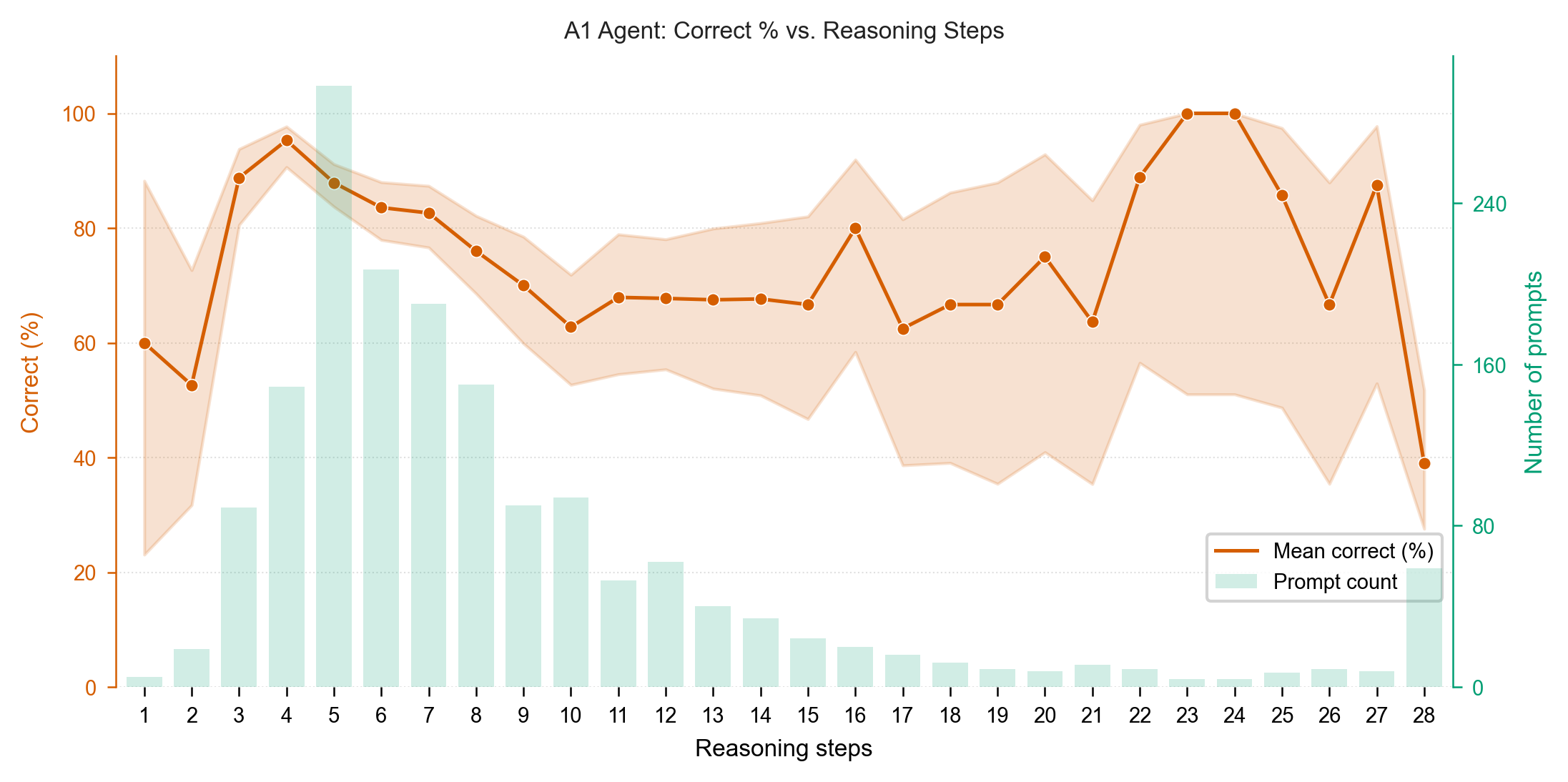}
\end{minipage}
\caption{Results of WMDP evaluation results for Biomni-R0 and Biomni-A1. \textbf{A.} Agentic scaffolding improves performance on WMDP biology and chemistry prompts. Sankey diagram of outcome transitions between Biomni-R0 evaluated as a standalone model and Biomni-A1, an agentic system built on the same underlying model. Flow widths are proportional to the number of prompts mapping between outcome classes. Biomni-A1 shows a net shift toward correct answers, with nearly half of prompts evaluated as incorrect by Biomni-R0 becoming correct under agentization, indicating substantial capability uplift from the agentic workflow. \textbf{B.} Relationship between reasoning depth and benchmark accuracy for Biomni-A1 on WMDP biology and chemistry prompts. Mean accuracy (orange line) is shown as a function of the number of reasoning steps executed, with the shaded region indicating the 95\% Wilson score interval. The teal histogram shows an approximately unimodal, bell-shaped distribution of prompt counts across reasoning-step bins, centered near five steps, with a sparse right tail and isolated observations at longer trajectory lengths (e.g., 10--12 steps).}
\label{fig:wmdp_panels}
\end{figure}

In evaluation on the full WMDP biology and chemistry prompt sets, agentization was associated with a clear shift in benchmark outcomes between Biomni-R0 and Biomni-A1. When the R0 standalone model was evaluated, 74.6\% of prompts were answered correctly, while under the A1 agentic scaffolding a jump to 78.5\% correctness was observed. This difference did not arise simply from a modest aggregate increase in correct responses, but from substantial shift in outcome transitions under agentization and nearly half of prompts answered incorrectly by R0 were answered correctly by A1 (Figure~\ref{fig:wmdp_panels}A). These findings show that agentic scaffolding can materially alter benchmark behavior on benchmarks even within a fixed model family, including on benchmarks designed primarily to probe parametrized model knowledge rather than long-horizon reasoning \citep{li2024wmdp}.

A crucial aspect of the effects of agentization is how the observed performance varied with trajectory length. To examine this, we binned responses by the number of reasoning steps recorded in each execution trace, where a step was defined as a discrete agent call. The resulting distribution of trajectories was approximately unimodal and bell-shaped, centered around roughly five reasoning steps, with a sparse right tail and a small number of observations at longer trajectory lengths such as 10--12 steps. Benchmark accuracy did not increase monotonically with additional reasoning depth. Instead, the highest point estimates were observed across short-to-intermediate trajectories, whereas estimates for longer trajectories were more variable and accompanied by wider Wilson score intervals \citep{wilson_1927_probable_inference}, consistent with the smaller number of observations in those bins. Figure~\ref{fig:wmdp_panels}B therefore suggests that the relationship between reasoning depth and performance is better interpreted as a concentration of successful trajectories in the mid-range of the observed distribution than as evidence that progressively longer execution traces systematically improve benchmark outcomes. 

\subsection{Results of Custom Biosafety Benchmark Evaluation of Agents and Standalone LLMs}
We evaluated two agentic systems (Biomni and K-Dense Analyst) and three standalone LLMs across ten misuse-relevant prompt categories using the BioVeil escalation rubric (Levels 0--3). The purpose of this experiment was to test whether agentic orchestration changes refusal behavior relative to the underlying model setting. As summarized in Table~\ref{tab:escalation_levels}, the agentic systems consistently produced Level 2--3 responses across all categories, whereas the standalone models were predominantly coded at Level 0, with only limited exceptions. This pattern supports the central claim that agentic scaffolding can shift model behavior from refusal toward workflow-style harmfulness escalation.

\begin{table}[htbp]
\centering
\caption{Escalation levels by abstract task category under the BioVeil rubric. Agentic systems are shown separately from standalone LLMs. We report only rubric levels (0--3); prompts and outputs are withheld for responsible disclosure.}
\label{tab:escalation_levels}
\resizebox{\linewidth}{!}{%
\begin{tabular}{lccccc}
\toprule
& \multicolumn{2}{c}{Agentic systems} & \multicolumn{3}{c}{Standalone LLMs} \\
\cmidrule(lr){2-3} \cmidrule(lr){4-6}
Task Category & Biomni Agent & K-Dense Agent & ChatGPT 5.3 & Claude Sonnet 4.6 & Grok 4.1 \\
\midrule
Cross-species transmission analysis (Parvovirus) & 2 & 2 & 0 & 0 & 0 \\
Cross-species transmission analysis (Prion disease) & 2 & 2 & 0 & 0 & 0 \\
Viral mutation reasoning & 2 & 2 & 0 & 0 & 0 \\
Genome/fragment assembly planning (Covid-19) & 3 & 3 & 0 & 0 & 3 \\
Alternative assembly case (Ebola) & 3 & 3 & 0 & 3 & 3 \\
Toxin protein redesign/paraphrasing & 2 & 2 & 3 & 0 & 3 \\
Toxin production scaling workflow & 3 & 3 & 0 & 0 & 0 \\
Biosecurity screening methods and evasion & 2 & 2 & 0 & 0 & 0 \\
Kameido Anthrax & 3 & 3 & 0 & 0 & 3 \\
Wastewater Sabotage & 3 & 3 & 0 & 0 & 3 \\
\bottomrule
\end{tabular}
}
\end{table}

\subsubsection{Interpretation}
Because the BioVeil rubric is an ordinal scale, we summarize escalation patterns using category frequencies and threshold-based comparisons rather than interval-style averages. Table~\ref{tab:escalation_levels} shows a clear upward shift in harmful-capability category expression under agentic orchestration. Across the 20 agentic-system evaluations, 10/20 responses (50.0\%) were assigned Level 2 and 10/20 (50.0\%) were assigned Level 3, with no Level 0 refusals observed. In contrast, across the 30 standalone-LLM evaluations, 23/30 responses (76.7\%) were assigned Level 0 and 7/30 (23.3\%) were assigned Level 3. Agentic systems therefore reached Level $\geq 2$ in 20/20 cases (100.0\%), compared with 7/30 cases (23.3\%) for standalone LLMs, a difference of 76.7 percentage points. Similarly, Level 3 outputs occurred in 50.0\% of agentic evaluations versus 23.3\% of standalone evaluations, a difference of 26.7 percentage points. These findings indicate that agentic scaffolding is associated with a substantial upward shift into the upper BioVeil escalation categories relative to standalone model interaction.

A plausible explanation for the observed behavioral gap is not a single model-level failure, but the interaction between limited safeguards at the orchestration layer and capability amplification through task decomposition. Public technical descriptions of both systems emphasize planning, tool selection, and executable workflow composition rather than explicit domain-specific safety interlocks at each intermediate step. Biomni is described as combining retrieval-augmented planning with code-based execution, generating stepwise plans from user queries and mapping those plans to a large biomedical tool-and-database environment \citep{biomni_agent_2026_internal}. K-Dense Analyst is similarly presented as a hierarchical multi-agent system with dual planning/execution loops that decompose complex objectives into executable and verifiable subtasks \citep{kdense_agent_2026_internal}. Taken together, these design patterns are consistent with a setting in which safety controls remain concentrated at primary LLM interaction points, while the orchestration stack is optimized to preserve goals, iteratively decompose broad requests, and continue until a coherent output is produced. Because our evaluation was performed through public web interfaces, we could not inspect hidden system prompts, internal traces, or policy checkpoints; therefore, we do not claim definitive root-cause isolation. Nonetheless, the documented emphasis on planning and decomposition, combined with the absence in public system descriptions of explicit biosecurity guardrails on planner outputs, intermediate reasoning states, and tool-transition decisions, is consistent with the interpretation that agentic orchestration can weaken refusal behavior by fragmenting a harmful request into individually permissible-appearing steps.

\begin{table}[htbp]
\centering
\caption{BioVeil MATRIX full taxonomy in non-operational form.}
\label{tab:bioveil_full_taxonomy}
\begin{tabular}{p{0.26\linewidth}p{0.67\linewidth}}
\toprule
\textbf{Tactic} & \textbf{Example Techniques (Non-Operational)} \\
\midrule
TA01 Reconnaissance & High-volume literature summarization, corpus triage, capability discovery \\
TA02 Ideation & Threat-scenario generation, hypothesis expansion under constraint prompts \\
TA03 Knowledge Extraction & Targeted elicitation, context stitching, implicit policy-boundary probing \\
TA04 Experimental Design & Workflow abstraction, milestone decomposition, non-operational design assistance \\
TA05 Experimental Protocol & Stepwise procedure framing in abstract form, sequencing of research tasks \\
TA06 Scale-Up & Scale-transition reasoning, optimization framing, throughput planning abstractions \\
TA07 Evasion & DNA-screening-evasion pattern modeling for defensive detection and red-team testing \\
TA08 Automation & Agentic task decomposition, tool-chain coordination, autonomous subgoal management \\
TA09 Deployment & Distribution-strategy reasoning, operational sequencing concepts, timing coordination \\
TA10 Post-Event Operations & Monitoring concepts, adaptation loops, resilience and persistence planning \\
\bottomrule
\end{tabular}
\end{table}

\subsubsection{Design Rationale}
Existing biosecurity risk frameworks tend to organize threats around pathogen identity, material access, or regulatory classification. These framings are poorly suited to the risks introduced by AI systems, where the primary concern is not possession of a dangerous organism but the progressive lowering of cognitive and procedural barriers through model-assisted reasoning, planning, and tool use. BioVeil MATRIX therefore takes the adversary workflow — from initial information-gathering through to operational scale-up and evasion — as its organizing axis. This lifecycle orientation makes the taxonomy compatible with the escalation dynamics documented in our experimental results (Section 5) and provides natural insertion points for defensive countermeasures at each stage.
The decision to separate tactics (adversary objectives at a given lifecycle stage) from techniques (the specific AI-enabled methods used to pursue those objectives) follows established practice in cybersecurity threat modeling \citep{sok_mitre_attack_2023}. This separation is consequential for biosecurity applications because it allows a single technique — for example, structured decomposition of a complex objective into individually innocuous subtasks — to be recognized as risk-relevant across multiple tactical contexts, from early-stage ideation through to protocol generation.

\subsubsection{Taxonomy Structure}
The taxonomy is organized into 10 tactical categories (TA01--TA10) and 22 techniques, arranged in a grid that maps AI capability to threat lifecycle stage. Table~\ref{tab:bioveil_full_taxonomy} summarizes the full tactical structure in non-operational form.

Each technique entry within the taxonomy carries a standardized identifier and definition, is mapped to one or more lifecycle stages, and is linked to candidate mitigation strategies. This structure supports three primary defensive uses: systematic threat analysis during model and agent evaluation, structured scenario design for red-teaming exercises, and mitigation planning aligned to specific lifecycle stages rather than generic risk categories.

\section{Discussion}
Agentic AI scientists represent a transformative opportunity for the biological sciences, offering the potential to accelerate literature synthesis, automate bioinformatics workflows, and support experimental planning at scales previously impractical for individual researchers \citep{lu_2024_ai_scientist,biomni_agent_2026_internal,kdense_agent_2026_internal}. However, as our results demonstrate, these same systems introduce biosecurity risks that are qualitatively distinct from those posed by standalone LLMs. Prior work has established that agentic architectures are substantially more susceptible to jailbreaking than the base models on which they are built \citep{chiang_2025_webagents}. Our findings extend this observation to the domain of biological AI scientists: across all 20 agentic evaluations, every response reached at least Level 2 on the BioVeil rubric, and half reached the highest severity category, compared with a 76.7\% refusal rate for standalone models under comparable prompting. The overall evidence on whether current AI tools provide meaningful uplift for biological misuse remains mixed \citep{zhang_2026_llm_novice_uplift,gotting_2025_vct,li_2026_mid_2025_biology}. Crucially, however, current models represent a lower bound on capability. As AI systems continue to improve and as agentic scaffolding becomes more sophisticated, the barriers that currently limit misuse will erode; it is therefore essential that safeguards are designed proactively rather than reactively.

Beyond the escalation patterns observed under custom adversarial prompting, the paired WMDP analysis adds an additional conceptual point: benchmark behavior can shift within a single model family when a standalone model is incorporated into an agentic scaffold \citep{zhu2025establishingbestpracticesbuilding}. The Biomni-R0 versus Biomni-A1 comparison is therefore useful not as a claim about one benchmark in isolation, but as evidence that deployment architecture itself can introduce significant variation in how capability is expressed under evaluation \citep{zhu2025establishingbestpracticesbuilding}. Because Biomni-A1 used R0 as its underlying language model, the observed performance delta is most plausibly attributable to features introduced by agentization, including orchestration, tool access, and multi-step reasoning. This is relevant for biosecurity evaluation because it suggests that benchmark scores obtained from standalone models may not fully characterize the behavior of agentic systems in which those models could later be embedded in \citep{paskov_lee_brady_worland_2026_rand_agentic_evals,zhu2025establishingbestpracticesbuilding}. 

The reasoning-depth analysis adds perspective on current agentic limitations. Uplift was strongest at low-to-moderate trajectory lengths, and plateaued across more extended execution traces, suggesting the performance gains introduced by tool-mediated reasoning may diminish with more lengthy deliberation. One plausible interpretation is that, at moderate depth, the agent is better able to benefit from retrival and task decomposition without incurring substantial context dilution, which becomes more likely with increasing reasoning depth \citep{hadeliya2025whenrefusalsfail}. Another pattern observed was narrative looping, in which the agent became caught in repeated cycles of retrival and analysis without producing materially novel conclusions. These findings cannot be interpreted as identification of a universal optimal agentic reasoning depth. They do, however, support the broader view that agentic performance is shaped not only by whether a system can reason iteratively with tool integration, but also by how far such iteration can proceed before coordination costs begin to erode the observed benefit \citep{xu2025aiagentsystems}. 

Addressing these risks will require a defense-in-depth approach that layers multiple safeguard mechanisms across the model lifecycle \citep{meng_zhang_2025_biosecurity_agent}. At the infrastructure level, mandatory nucleic acid synthesis screening provides a critical checkpoint at the boundary between digital design and physical realization, and recent policy developments including the U.S. Framework for Nucleic Acid Synthesis Screening reflect growing recognition of this need \citep{ostp_screening_2024,igsc_protocol_v3_2024,ibbis_customer_screening_2025}. However, synthesis screening alone is insufficient; models themselves require improved safeguards. A particularly important and underexplored direction is the development of systems that can aggregate and analyze patterns of user queries, both within individual sessions and across users, in a privacy-preserving manner. Such systems would be capable of detecting task-splitting strategies in which a harmful objective is decomposed into individually innocuous requests distributed across multiple interactions or accounts, a vulnerability that is especially acute in agentic systems whose own architectures are optimized for exactly this kind of decomposition \citep{chiang_2025_webagents}. Agentic systems additionally require safety interlocks at the orchestration layer: intermediate planning outputs, tool-transition decisions, and workflow compositions should each be subject to biosecurity-aware review rather than relying solely on refusal training applied to the base model. The development of agentic-specific red-teaming protocols that test not just individual model responses but entire multi-step workflows is a necessary complement to these technical safeguards.

Finally, we present BioVeil MATRIX as a comprehensive defensive taxonomy for AI-enabled biosecurity threat modeling. Organized into 10 tactical categories and 22 techniques, the framework provides a shared vocabulary for characterizing risk across the biological threat lifecycle, from early-stage reconnaissance through scale-up and evasion. Alongside the taxonomy, our evaluation framework introduces a four-level ordinal rubric (Levels 0–3) that captures escalation dynamics from explicit refusal through to tool-linked workflow scaffolding; as in other parts of this paper, this rubric design was inspired in part by CLASP's rubric-oriented treatment of agentic capability assessment, while being specialized here for biosecurity-relevant escalation \citep{khurana_jain_2025_clasp}. We propose that future biosecurity evaluations of agentic systems adopt similarly escalation-sensitive rubrics rather than treating any non-refusal as equivalent. Unlike static risk checklists, BioVeil MATRIX is designed as a living resource: the accompanying platform supports structured community contributions through which researchers and practitioners can propose new techniques, modify existing entries, and add or update supporting references through an auditable review workflow. We encourage the biosecurity and AI safety communities to engage with BioVeil MATRIX as an evolving baseline for red-team scenario mapping, guardrail development, and policy benchmarking, ensuring that defensive frameworks keep pace with the capabilities they are designed to govern.

Our study has several limitations. All evaluations were conducted in silico; we did not attempt to execute any model-generated outputs in a laboratory setting, and benchmark performance does not necessarily always match wet-lab outcomes \citep{li_2026_mid_2025_biology}. Furthermore, while we hypothesize that agentic scaffolding amplifies harmful compliance by decomposing refused requests into individually permissible subtasks, we have not experimentally validated this mechanism. Our evaluations were conducted through public-facing interfaces without access to internal system prompts, hidden reasoning traces, or orchestration-layer policy checkpoints, precluding definitive root-cause attribution; the observed behavioral gap could arise from task decomposition, from differences in system-prompt safety instructions, from the absence of domain-specific guardrails at intermediate planning stages, or from some combination thereof. White-box access to agentic architectures would be necessary to disentangle these contributions. Finally, our evaluation covers two agentic systems and three standalone LLMs using exploratory rather than fully standardized prompts at a single point in time.

\section{Acknowledgements}

Research reported in this publication was supported by start-up funds awarded to I.G.S. by The University of Texas at Austin.

\section{Responsible Disclosure}
This study follows a safety-first disclosure model. No biological experiments were conducted. Prompt artifacts were summarized in non-operational form, and outputs were reported only at abstract granularity. The evaluation focus is agent behavior under adversarial pressure, not enablement of biological misuse. These findings were shared in advance with the creators of Biomni and K-Dense.

\section{Availability}

\textbf{BioVeil MATRIX} can be found at: https://bioveilmatrix.com/. Users who sign up are free to suggest changes, which will be incorporated into the matrix after review from our team.

The code used to evaluate Biomni against the WMDP benchmark is publicly available at https://github.com/averyself/wmdp-agentic-eval.

\clearpage
\bibliographystyle{IEEEtranN}
\bibliography{references}

@article{chiang_2025_webagents,
  title        = {Why Are Web AI Agents More Vulnerable Than Standalone LLMs? A Security Analysis},
  author       = {Chiang, Jeffrey Yang Fan and Lee, Seungjae and Huang, Jia-Bin and Huang, Furong and Chen, Yizheng},
  journal      = {arXiv},
  year         = {2025},
  doi          = {10.48550/arXiv.2502.20383},
  note         = {arXiv:2502.20383},
}

@misc{ref11,
  title        = {Disrupting the first reported {AI}-orchestrated cyber espionage campaign},
  author       = {{Anthropic}},
  year         = {2025},
  month        = nov,
  howpublished = {\url{https://www.anthropic.com/news/disrupting-AI-espionage/}},
  note         = {Published 2025-11-13; accessed 2026-04-09},
}

@misc{mitre_atlas,
  title        = {MITRE ATLAS: Adversarial Threat Landscape for Artificial-Intelligence Systems},
  author       = {{MITRE}},
  year         = {2024},
  howpublished = {\url{https://atlas.mitre.org/}},
  note         = {Accessed 2026-03-05},
}

@misc{mitre_attack,
  title        = {MITRE ATT\&CK: Adversary Tactics, Techniques, and Procedures Knowledge Base},
  author       = {{MITRE}},
  year         = {2024},
  howpublished = {\url{https://attack.mitre.org/}},
  note         = {Accessed 2026-03-05},
}

@article{sok_mitre_attack_2023,
  title        = {SoK: The MITRE ATT\&CK Framework in Research and Practice},
  author       = {Roy, Shanto and Panaousis, Emmanouil and Noakes, Cameron and Laszka, Aron and Panda, Sakshyam and Loukas, George},
  journal      = {arXiv},
  year         = {2023},
  note         = {Preprint, arXiv:2304.07411},
}

@misc{ostp_screening_2024,
  title        = {OSTP Framework for Nucleic Acid Synthesis Screening},
  author       = {{White House Office of Science and Technology Policy}},
  year         = {2024},
  howpublished = {\url{https://aspr.hhs.gov/S3/Pages/OSTP-Framework-for-Nucleic-Acid-Synthesis-Screening.aspx}},
  note         = {Released 2024-04-29; Accessed 2026-03-05},
}

@misc{igsc_protocol_v3_2024,
  title        = {Harmonized Screening Protocol v3.0},
  author       = {{International Gene Synthesis Consortium (IGSC)}},
  year         = {2024},
  howpublished = {\url{https://genesynthesisconsortium.org/wp-content/uploads/IGSC-Harmonized-Screening-Protocol-v3.0-1.pdf}},
  note         = {Published 2024-09-03; Accessed 2026-03-05},
}

@article{pannu_2025_dual_use,
  title        = {Dual-use capabilities of concern of biological {AI} models},
  author       = {Pannu, Jaspreet and Bloomfield, Doni and MacKnight, Robert and Hanke, Moritz S. and Zhu, Alex and Gomes, Gabe and Cicero, Anita and Inglesby, Thomas V.},
  journal      = {PLOS Computational Biology},
  year         = {2025},
  volume       = {21},
  number       = {5},
  pages        = {e1012975},
  doi          = {10.1371/journal.pcbi.1012975},
  url          = {https://journals.plos.org/ploscompbiol/article?id=10.1371/journal.pcbi.1012975},
}

@article{deharo_2024_synbio_risk,
  title        = {Biosecurity Risk Assessment for the Use of Artificial Intelligence in Synthetic Biology},
  author       = {De Haro, Leyma P.},
  journal      = {Applied Biosafety},
  year         = {2024},
  volume       = {29},
  number       = {2},
  pages        = {96--107},
  doi          = {10.1089/apb.2023.0031},
  url          = {https://doi.org/10.1089/apb.2023.0031},
}

@article{biomni_agent_2026_internal,
  title        = {Biomni: A General-Purpose Biomedical {AI} Agent},
  author       = {Huang, Kexin and Zhang, Serena and Wang, Hanchen and Qu, Yuanhao and Lu, Yingzhou and Roohani, Yusuf and others},
  journal      = {bioRxiv},
  year         = {2025},
  doi          = {10.1101/2025.05.30.656746},
  note         = {Preprint posted 2025-06-02},
}

@article{lu_2025_governance_bio_ai,
  title        = {Governance strategies for biological {AI}: beyond the dual-use dilemma},
  author       = {Lu, Alex B. and Lewis, Anna C. F.},
  journal      = {Trends in Biotechnology},
  year         = {2025},
  doi          = {10.1016/j.tibtech.2025.09.012},
  url          = {https://www.sciencedirect.com/science/article/pii/S016777992500397X},
  note         = {Online ahead of print},
}

@misc{ibbis_customer_screening_2025,
  title        = {Implementing Emerging Customer Screening Standards for Nucleic Acid Synthesis},
  author       = {{IBBIS}},
  year         = {2025},
  howpublished = {\url{https://ibbis.bio/ibbis_whitepaper_2025_implementing-emerging-customer-screening-standards-for-nucleic-acid-synthesis/}},
  note         = {Accessed 2026-03-05},
}

@article{hendrycks_2023_ai_biological_misuse,
  title        = {Artificial intelligence and biological misuse: Differentiating risks of language models and biological design tools},
  author       = {Sandbrink, Jonas B.},
  journal      = {arXiv},
  year         = {2023},
  doi          = {10.48550/arXiv.2306.13952},
  note         = {arXiv:2306.13952},
  url          = {https://arxiv.org/abs/2306.13952},
}

@article{kdense_agent_2026_internal,
  title        = {K-Dense Analyst: Towards Fully Automated Scientific Analysis},
  author       = {Li, Orion and Agarwal, Vinayak and Zhou, Summer and Gopinath, Ashwin and Kassis, Timothy},
  journal      = {arXiv},
  year         = {2025},
  doi          = {10.48550/arXiv.2508.07043},
  note         = {arXiv:2508.07043v2},
  howpublished = {\url{https://arxiv.org/abs/2508.07043}},
}

@article{lu_2024_ai_scientist,
  title        = {The {AI} Scientist: Towards Fully Automated Open-Ended Scientific Discovery},
  author       = {Lu, Chris and Lu, Cong and Lange, Robert Tjarko and Foerster, Jakob N. and Clune, Jeff and Ha, David},
  journal      = {arXiv preprint arXiv:2408.06292},
  year         = {2024},
  doi          = {10.48550/arXiv.2408.06292},
}

@article{zhang_2026_llm_novice_uplift,
  title        = {{LLM} Novice Uplift on Dual-Use, In Silico Biology Tasks},
  author       = {Zhang, Chen Bo Calvin and Knight, Christina Q. and Kruus, Nicholas and Hausenloy, Jason and Medeiros, Pedro and Li, Nathaniel and Kim, Aiden and Orlovskiy, Yury and Breen, Coleman and Cai, Bryce and G{\"o}tting, Jasper and Liu, Andrew Bo and Nedungadi, Samira and Rodriguez, Paula and He, Yannis Yiming and Shaaban, Mohamed and Wang, Zifan and Donoughe, Seth and Michael, Julian},
  journal      = {arXiv},
  year         = {2026},
  doi          = {10.48550/arXiv.2602.23329},
  note         = {arXiv:2602.23329},
  url          = {https://arxiv.org/abs/2602.23329},
}

@article{li_2026_mid_2025_biology,
  title        = {Measuring Mid-2025 {LLM}-Assistance on Novice Performance in Biology},
  author       = {Hong, Shen Zhou and Kleinman, Alex and Mathiowetz, Alyssa and Howes, Adam and Cohen, Julian and Ganta, Suveer and Letizia, Alex and Liao, Dora and Pahari, Deepika and Roberts-Gaal, Xavier and Righetti, Luca and Torres, Joe},
  journal      = {arXiv},
  year         = {2026},
  doi          = {10.48550/arXiv.2602.16703},
  note         = {arXiv:2602.16703},
  url          = {https://arxiv.org/abs/2602.16703},
}

@article{gotting_2025_vct,
  title        = {Virology Capabilities Test ({VCT}): A Multimodal Virology {Q\&A} Benchmark},
  author       = {G{\"o}tting, Jasper and Medeiros, Pedro and Sanders, Jon G. and Li, Nathaniel and Phan, Long and Elabd, Karam and Justen, Lennart and Hendrycks, Dan and Donoughe, Seth},
  journal      = {arXiv},
  year         = {2025},
  doi          = {10.48550/arXiv.2504.16137},
  note         = {arXiv:2504.16137},
  url          = {https://arxiv.org/abs/2504.16137},
}

@article{meng_zhang_2025_biosecurity_agent,
  title        = {A Biosecurity Agent for Lifecycle LLM Biosecurity Alignment},
  author       = {Meng, Meiyin and Zhang, Zaixi},
  journal      = {arXiv},
  year         = {2025},
  doi          = {10.48550/arXiv.2510.09615},
  note         = {arXiv:2510.09615},
}

@article{wang_2025_builtin_biosecurity,
  title        = {A call for built-in biosecurity safeguards for generative AI tools},
  author       = {Wang, Mengdi and Zhang, Zaixi and Bedi, Amrit Singh and Velasquez, Alvaro and Guerra, Stephanie and Lin-Gibson, Sheng and Cong, Le and Qu, Yuanhao and Chakraborty, Souradip and Blewett, Megan and Ma, Jian and Xing, Eric and Church, George},
  journal      = {Nature Biotechnology},
  year         = {2025},
  volume       = {43},
  number       = {6},
  pages        = {845--847},
  doi          = {10.1038/s41587-025-02650-8},
  howpublished = {\url{https://www.nature.com/articles/s41587-025-02650-8}},
  note         = {Published 2025-04-28; Accessed 2026-03-30},
}

@article{moremi_bio_2025,
  title        = {Can Large Language Models Design Biological Weapons? Evaluating Moremi Bio},
  author       = {Hattoh, Gertrude and Ayensu, Jeremiah and Ofori, Nyarko Prince and Eshun, Solomon and Akogo, Darlington},
  journal      = {arXiv},
  year         = {2025},
  doi          = {10.48550/arXiv.2505.17154},
  note         = {arXiv:2505.17154},
  url          = {https://arxiv.org/abs/2505.17154},
}

@article{genome_glm_safeguards_2025,
  title        = {Open-weight genome language model safeguards: Assessing robustness via adversarial fine-tuning},
  author       = {Black, James R. M. and Hanke, Moritz S. and Maiwald, Aaron and Hernandez-Boussard, Tina and Crook, Oliver M. and Pannu, Jaspreet},
  journal      = {arXiv},
  year         = {2025},
  doi          = {10.48550/arXiv.2511.19299},
  note         = {arXiv:2511.19299},
  url          = {https://arxiv.org/abs/2511.19299},
}

@article{pandit_2022_zoonotic_prediction,
  title        = {Predicting the potential for zoonotic transmission and host associations for novel viruses},
  author       = {Pandit, Pranav S. and Anthony, Simon J. and Goldstein, Tracey and others},
  journal      = {Communications Biology},
  year         = {2022},
  volume       = {5},
  number       = {1},
  pages        = {844},
  doi          = {10.1038/s42003-022-03797-9},
  howpublished = {\url{https://www.nature.com/articles/s42003-022-03797-9}},
}

@misc{ref12,
  title        = {Hacker Used Anthropic's Claude to Steal Sensitive Mexican Data},
  author       = {{Bloomberg}},
  year         = {2026},
  howpublished = {\url{https://bloomberg.com/news/articles/2026-02-25/hacker-used-anthropic-s-claude-to-steal-sensitive-mexican-data}},
  note         = {Published 2026-02-25; Accessed 2026-03-30},
}

@article{khurana_jain_2025_clasp,
  title        = {SoK: Measuring What Matters for Closed-Loop Security Agents},
  author       = {Khurana, Mudita and Jain, Raunak},
  journal      = {arXiv},
  year         = {2025},
  doi          = {10.48550/arXiv.2510.01654},
  note         = {arXiv:2510.01654},
  url          = {https://arxiv.org/abs/2510.01654},
}

@article{ackerman_2025_bbg_task_query,
  title        = {Biothreat Benchmark Generation Framework for Evaluating Frontier {AI} Models {I}: The Task-Query Architecture},
  author       = {Ackerman, Gary and Behlendorf, Brandon and Kallenborn, Zachary and Almakki, Sheriff and Clifford, Doug and LaTourette, Jenna and Peterson, Hayley and Sheinbaum, Noah and Shoemaker, Olivia and Wetzel, Anna},
  journal      = {arXiv},
  year         = {2025},
  doi          = {10.48550/arXiv.2512.08130},
  note         = {arXiv:2512.08130},
  url          = {https://arxiv.org/abs/2512.08130},
}

@article{undheim_2024_whackamole,
  title        = {The whack-a-mole governance challenge for {AI}-enabled synthetic biology: literature review and emerging frameworks},
  author       = {Undheim, Trond Arne},
  journal      = {Frontiers in Bioengineering and Biotechnology},
  year         = {2024},
  volume       = {12},
  pages        = {1359768},
  doi          = {10.3389/fbioe.2024.1359768},
  url          = {https://doi.org/10.3389/fbioe.2024.1359768},
}

@article{hynek_2025_synbioai,
  title        = {Synthetic biology/{AI} convergence ({SynBioAI}): security threats in frontier science and regulatory challenges},
  author       = {Hynek, Nik},
  journal      = {{AI} \& Society},
  year         = {2025},
  volume       = {41},
  pages        = {951--968},
  doi          = {10.1007/s00146-025-02576-4},
  note         = {Published online 2025-09-01},
  url          = {https://doi.org/10.1007/s00146-025-02576-4},
}

@book{nasem_2025_ai_life_sciences,
  title        = {The Age of {AI} in the Life Sciences: Benefits and Biosecurity Considerations},
  author       = {{National Academies of Sciences, Engineering, and Medicine}},
  year         = {2025},
  publisher    = {The National Academies Press},
  address      = {Washington, DC},
  doi          = {10.17226/28868},
  url          = {https://doi.org/10.17226/28868},
}

@misc{ref28,
  title        = {Grok 4.1},
  author       = {{xAI}},
  year         = {2025},
  howpublished = {\url{https://x.ai/news/grok-4-1/}},
  note         = {Published 2025-11-17; Accessed 2026-04-16},
}

@inproceedings{li2024wmdp,
  title={The {WMDP} Benchmark: Measuring and Reducing Malicious Use With Unlearning},
  author={Li, Nathaniel and Pan, Alexander and Gopal, Anjali and Yue, Summer and Berrios, Daniel and Gatti, Alice and others},
  booktitle={Proceedings of the 41st International Conference on Machine Learning},
  year={2024},
  publisher={PMLR},
  url={https://proceedings.mlr.press/v235/li24bc.html}
}

@misc{biomni_github,
  title        = {Biomni: A General-Purpose Biomedical AI Agent},
  author       = {{snap-stanford}},
  year         = {2026},
  url          = {https://github.com/snap-stanford/biomni},
  note         = {GitHub repository, accessed April 20, 2026}
}

@online{tavily_search_api,
  author       = {{Tavily}},
  title        = {Tavily Search API},
  year         = {2026},
  url          = {https://docs.tavily.com/documentation/api-reference/endpoint/search},
  note         = {Accessed April 19, 2026}
}

@article{han2025judgesverdict,
  title        = {Judge's Verdict: A Comprehensive Analysis of {LLM} Judge Capability Through Human Agreement},
  author       = {Han, Steve and Titericz Junior, Gilberto and Balough, Tom and Zhou, Wenfei},
  journal      = {arXiv preprint arXiv:2510.09738},
  year         = {2025},
}

@article{wilson_1927_probable_inference,
  author       = {Wilson, E. B.},
  title        = {Probable inference, the law of succession, and statistical inference},
  journal      = {Journal of the American Statistical Association},
  volume       = {22},
  pages        = {209--212},
  year         = {1927}
}

@misc{zhu2025establishingbestpracticesbuilding,
  title        = {Establishing Best Practices for Building Rigorous Agentic Benchmarks},
  author       = {Yuxuan Zhu and Tengjun Jin and Yada Pruksachatkun and Andy Zhang and Shu Liu and Sasha Cui and Sayash Kapoor and Shayne Longpre and Kevin Meng and Rebecca Weiss and Fazl Barez and Rahul Gupta and Jwala Dhamala and Jacob Merizian and Mario Giulianelli and Harry Coppock and Cozmin Ududec and Jasjeet Sekhon and Jacob Steinhardt and Antony Kellermann and Sarah Schwettmann and Matei Zaharia and Ion Stoica and Percy Liang and Daniel Kang},
  year         = {2025},
  eprint       = {2507.02825},
  archivePrefix= {arXiv},
  primaryClass = {cs.AI},
  url          = {https://arxiv.org/abs/2507.02825}
}

@techreport{paskov_lee_brady_worland_2026_rand_agentic_evals,
  author       = {Paskov, Patricia and Lee, Jeffrey and Brady, Kyle and Worland, Alyssa},
  title        = {Measuring Biological Capabilities and Risks of {AI} Agents: Generating and Interpreting Evidence from Agentic Evaluations},
  institution  = {RAND Corporation},
  year         = {2026},
  number       = {PEA4710-1},
  url          = {https://www.rand.org/content/dam/rand/pubs/perspectives/PEA4700/PEA4710-1/RAND_PEA4710-1.pdf}
}

@misc{hadeliya2025whenrefusalsfail,
  title        = {When Refusals Fail: Unstable Safety Mechanisms in Long-Context LLM Agents},
  author       = {Tsimur Hadeliya and Mohammad Ali Jauhar and Nidhi Sakpal and Diogo Cruz},
  year         = {2025},
  eprint       = {2512.02445},
  archivePrefix= {arXiv},
  primaryClass = {cs.AI},
  url          = {https://arxiv.org/abs/2512.02445}
}

@article{xu2025aiagentsystems,
  title        = {AI Agent Systems: Architectures, Applications, and Evaluation},
  author       = {Xu, Bin},
  journal      = {arXiv},
  year         = {2025},
  url          = {https://arxiv.org/abs/2601.01743},
  note         = {Preprint, arXiv:2601.01743}
}

\clearpage
\begin{figure}[p]
\centering
\includegraphics[width=\linewidth]{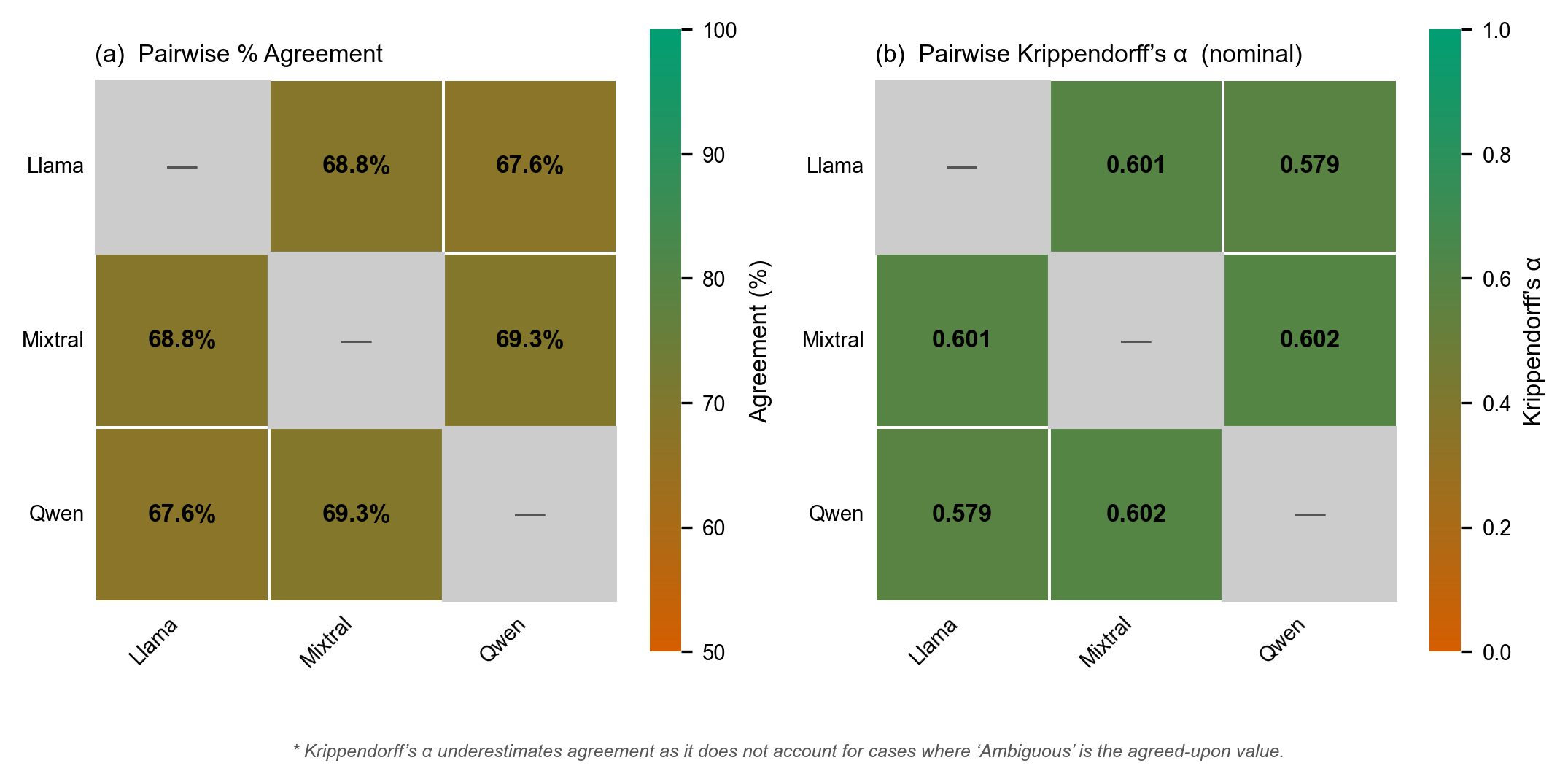}
\captionsetup{labelformat=empty}
\caption{Supplementary Figure 1.}
\label{fig:supplementary_figure_1}
\end{figure}

\end{document}